\documentclass[sigconf]{acmart}
\AtBeginDocument{%
  }

\setcopyright{acmlicensed}
\copyrightyear{2018}
\acmYear{2018}
\acmDOI{XXXXXXX.XXXXXXX}
\acmConference[Conference acronym 'XX]{Make sure to enter the correct
  conference title from your rights confirmation email}{June 03--05,
  2018}{Woodstock, NY}
\acmISBN{978-1-4503-XXXX-X/2018/06}

\usepackage{booktabs}
\usepackage{multirow}
\usepackage{colortbl}
\usepackage{xcolor}
\usepackage{amsmath, amssymb}
\usepackage{algorithm}
\usepackage{algpseudocode}
\usepackage{enumitem}
\usepackage{bm}
\usepackage{soul}
\usepackage{afterpage} 
\usepackage{amssymb}
\usepackage{pdflscape}

\copyrightyear{2025}
\acmYear{2025}
\setcopyright{cc}
\setcctype{by-nc-sa}
\acmConference[CIKM '25]{Proceedings of the 34th ACM International Conference on Information and Knowledge Management}{November 10--14, 2025}{Seoul, Republic of Korea}
\acmBooktitle{Proceedings of the 34th ACM International Conference on Information and Knowledge Management (CIKM '25), November 10--14, 2025, Seoul, Republic of Korea}\acmDOI{10.1145/3746252.3761572}
\acmISBN{979-8-4007-2040-6/2025/11}

\begin{document}

\title{DeepAries: Adaptive Rebalancing Interval Selection for Enhanced Portfolio Selection}


\author{Jinkyu Kim}
\orcid{0000-0001-9484-7541}
\affiliation{%
  \institution{Korea University}
  \city{Seoul}
  \country{South Korea}
}
\email{no100kill@korea.ac.kr}

\author{Hyunjung Yi}
\orcid{0009-0000-6617-3105}
\affiliation{%
  \institution{Korea University}
  \city{Seoul}
  \country{South Korea}
}
\email{ruby3672@korea.ac.kr}

\author{Mogan Gim}
\orcid{0000-0002-6458-7723}
\affiliation{%
  \institution{Hankuk University of Foreign Studies}
  \city{Yongin}
  \country{South Korea}
}
\email{gimmogan@hufs.ac.kr}

\author{Donghee Choi}
\authornote{Corresponding authors.}
\orcid{0000-0002-8857-9680}
\affiliation{%
  \institution{Pusan National University}
  \city{Busan}
  \country{South Korea}
}
\affiliation{%
  \institution{Imperial College London}
  \city{London}
  \country{United Kingdom}
}
\email{dchoi@pusan.ac.kr}

\author{Jaewoo Kang}
\authornotemark[1] 
\orcid{0000-0001-6798-9106}
\affiliation{%
  \institution{Korea University}
  \city{Seoul}
  \country{South Korea}
}
\email{kangj@korea.ac.kr}

\renewcommand{\shortauthors}{Jinkyu Kim, Hyunjung Yi, Mogan Gim, Donghee Choi, \& Jaewoo Kang}

\definecolor{mydarkgreen}{RGB}{0,80,0} 
\newcommand{\mg}[1]{{\color{mydarkgreen}\textbf{[MG:} #1\textbf{]}}}
\newcommand{\hj}[1]{{\color{orange}\textbf{[HJ:} #1\textbf{]}}}

\newcommand{\modelname}{\textsc{DeepAries}}
\newcommand{\projectsite}{\url{https://deep-aries.github.io/}}
\newcommand{\githubrepo}{\url{https://github.com/dmis-lab/DeepAries}}

\begin{abstract}
We propose \modelname, a novel deep reinforcement learning framework for dynamic portfolio management that jointly optimizes the timing and allocation of rebalancing decisions. Unlike prior reinforcement learning methods that employ fixed rebalancing intervals regardless of market conditions, \modelname~ adaptively selects optimal rebalancing intervals along with portfolio weights to reduce unnecessary transaction costs and maximize risk-adjusted returns. Our framework integrates a Transformer-based state encoder, which effectively captures complex long-term market dependencies, with Proximal Policy Optimization (PPO) to generate simultaneous discrete (rebalancing intervals) and continuous (asset allocations) actions. Extensive experiments on multiple real-world financial markets demonstrate that \modelname~ significantly outperforms traditional fixed-frequency and full-rebalancing strategies in terms of risk-adjusted returns, transaction costs, and drawdowns. Additionally, we provide a live demo of \modelname~at~\projectsite, along with the source code and dataset at~\githubrepo, illustrating \modelname' capability to produce interpretable rebalancing and allocation decisions aligned with shifting market regimes. Overall, \modelname~ introduces an innovative paradigm for adaptive and practical portfolio management by integrating both timing and allocation into a unified decision-making process.
\end{abstract}

\ccsdesc[500]{Information systems~Expert systems}
\ccsdesc[500]{Information systems~Data mining}
\ccsdesc[500]{Computing methodologies~Artificial intelligence}
\ccsdesc[500]{Computing methodologies~Learning paradigms}



\keywords{Portfolio Management; Adaptive Rebalancing Interval Selection; Artificial Intelligence in Finance; Deep Reinforcement Learning}

\begin{teaserfigure}
  \includegraphics[width=\textwidth]{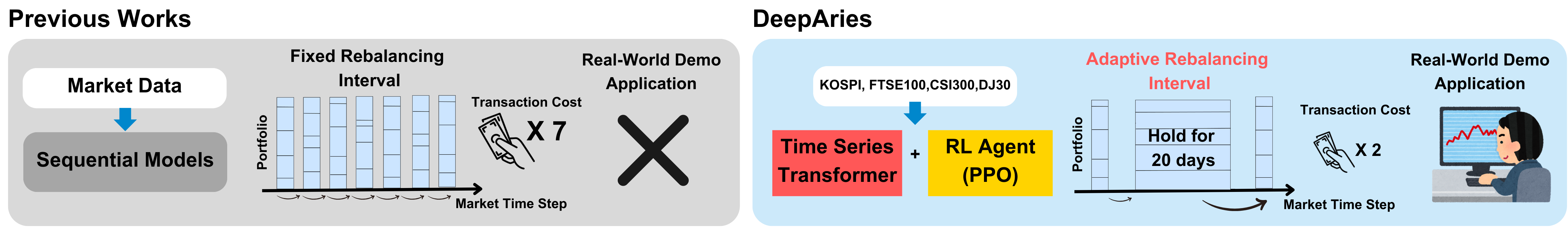}
  \caption{Illustrative comparison between previously introduced portfolio selection models and \modelname}
  \Description{Overview of \modelname.}
  \label{fig:teaser}
\end{teaserfigure}


\maketitle

\section{Introduction}

Portfolio optimization traditionally aims to balance risk and return via asset allocation strategies. Classical methods, such as Markowitz’s mean-variance optimization ~\cite{markowitz1952portfolio} and Sharpe’s Capital Asset Pricing Model (CAPM) \cite{sharpe1964capital}, focus \textbf{predominantly on static or single-period decisions}. However, real-world portfolio management demands dynamic and sequential decision-making, specifically addressing two interconnected challenges: \textit{when} to rebalance and \textit{how} to allocate.

While recent reinforcement learning (RL) approaches have successfully tackled dynamic allocation problems, existing frameworks (e.g., PGPortfolio \cite{jiang2017deep}, FinRL \cite{Liu2020}, DeepTrader \cite{wang2021deeptrader}) typically employ fixed rebalancing intervals and fully liquidate portfolios at every decision step. Such practices ignore adaptive market conditions and incur significant, often unnecessary, transaction costs. For instance, frequent full rebalancing during stable markets generates excessive transaction fees, whereas infrequent rebalancing can delay necessary adjustments during market volatility\cite{daryanani2008opportunistic,perold1988dynamic}. This highlights a critical gap: current methods fail to intelligently determine optimal rebalancing timing.

\modelname~addresses a novel hybrid decision-making problem, newly framed in this work, which combines discrete rebalancing interval selection with continuous asset allocation.
The discrete nature of rebalancing stems from fixed, actionable time points, while allocation decisions inherently require smooth adjustments to asset weights.
To tackle this, we propose an RL framework explicitly designed to adaptively determine rebalancing intervals and asset allocations by capturing evolving market conditions.
Specifically, \modelname~leverages a Transformer-based encoder to extract rich \textbf{temporal and cross-asset market dynamics} and utilizes \textbf{Proximal Policy Optimization (PPO)} to train an agent capable of \textbf{simultaneously choosing discrete rebalancing intervals and continuous asset allocations}. 
By adaptively determining whether to rebalance immediately or hold current positions based on market signals, \textbf{\modelname~reduces unnecessary turnover and transaction costs} while effectively capturing opportunities during market shifts.

Empirical results show that \modelname~achieves state-of-the-art performance across four international markets, improving both returns and risk-adjusted returns compared to existing portfolio selection methods.
Through comprehensive ablation studies, we validate the effectiveness of our adaptive rebalancing strategy and the iTransformer\cite{liu2023itransformer} backbone, both of which contribute significantly to the overall performance.

Our contributions are as follows:
\begin{itemize}
    \item We introduce a \emph{Adaptive Rebalancing Interval Selection} RL formulation for portfolio management, extending beyond the common practice of fixed trading intervals.
    \item We integrate a Transformer-based encoder to process multi-asset sequential data, capturing both temporal dependencies and cross-asset relationships in a unified latent space. Formulating a tractable joint action space optimized using PPO, balancing interval choices and asset allocation simultaneously.
    \item We demonstrate empirically, across multiple equity markets, that adaptively selecting the rebalancing intervals leads to \emph{improved risk-adjusted returns} and reduced transaction costs compared to fixed, daily rebalancing daily strategies.
    \item We validate the real-world applicability of our approach through an interactive demonstration, highlighting its potential to empower investors with timely portfolio updates and more informed decision-making.
\end{itemize}

\section{Related Works}
\label{sec:related}

\subsection{Classical and Static Portfolio Optimization}
Modern portfolio theory, introduced by Markowitz~\cite{markowitz1952portfolio}, established a static framework based on mean-variance optimization. Such traditional methods, including the Capital Asset Pricing Model (CAPM)\cite{sharpe1964capital}, primarily focus on single-period decisions and fail to dynamically adapt to evolving market conditions. Consequently, they often rely on heuristic periodic rebalancing strategies, which may either respond inadequately to rapid market shifts or incur unnecessary transaction costs during stable market conditions\cite{benhamou2023,OlivaresNadal2018}.

\subsection{Reinforcement Learning for Portfolio Management}
Recent RL approaches have significantly advanced portfolio management by dynamically adjusting asset allocations. Jiang et al.\cite{jiang2017deep} introduced the Ensemble of Identical Independent Evaluators (EIIE), utilizing CNN-based networks for asset-wise evaluation. Extensions to this approach include cost-sensitive frameworks proposed by Zhang et al.\cite{Zhang2022}, explicitly incorporating transaction costs and risk penalties into the reward function~\cite{Almahdi2017,Yan2024,HWANG2023104285}. Following this line of research, Kim et al.~\cite{kim2023hadaps} proposed HADAPS, a model applying an Asynchronous Advantage Actor-Critic algorithm to the portfolio selection problem.

\subsection{Transformer-Based Time Series Models}
Transformer architectures have proven highly effective in financial applications~\cite{Li2023}, capturing complex temporal dependencies without extensive manual feature engineering. Xu et al.\cite{Xu2020} introduced a Relation-Aware Transformer (RAT), significantly improving performance by modeling asset interactions through self-attention mechanisms. Similarly, Choi et al.\cite{Choi_2024} proposed DeepClair, integrating transformer-based market forecasts into an RL framework, thereby enhancing adaptive decision-making capabilities, especially in volatile market conditions.

\subsection{Limitations of Existing RL Approaches and Adaptive Rebalancing Interval Selection}
Most existing RL-based portfolio optimization methods, such as EIIE~\cite{jiang2017deep}, FinRL~\cite{Liu2020}, DeepTrader~\cite{wang2021deeptrader},HADAPS~\cite{kim2023hadaps} and DeepClair ~\cite{Choi_2024} employ full rebalancing at fixed, often daily intervals. Each step entails fully liquidating existing holdings before repurchasing new allocations, increasing transaction costs, especially during stable market periods where adjustments might be minimal. Alternatives involving partial adjustments are underexplored due to the complexity involved in dynamically determining the appropriate timing and scale of rebalancing.

To address these limitations, we introduce \textit{Adaptive Rebalancing Interval Selection} within our \modelname~framework. Unlike previous methods, \modelname~ dynamically selects rebalancing intervals based on real-time market conditions, significantly reducing unnecessary transaction costs. Combining a Transformer-based encoder with Proximal Policy Optimization (PPO), \modelname~effectively manages the simultaneous selection of rebalancing intervals and portfolio allocations, achieving an optimal balance between responsiveness and cost-efficiency.

\section{Problem Formulation}
\label{sec:problem}

We formulate adaptive portfolio management as an RL problem, where the agent learns to dynamically determine both \emph{when to rebalance} the portfolio and \emph{how to allocate assets} at each rebalancing decision point.

Let \(\mathcal{X} = \{x_1, x_2, \dots, x_N\}\) denote a universe of \(N\) tradable assets. At each decision epoch \(T_m\), the agent observes the market state as a tensor:
\[
\mathbf{s}(T_m) \in \mathbb{R}^{N \times \tau \times F},
\]
where \(\tau\) is the look-back window length and \(F\) is the number of features (e.g., open, high, low, close prices) per asset. Thus, \(\mathbf{s}(T_m)\) represents historical price features for all assets over the past \(\tau\) timesteps.

Based on the observed state \(\mathbf{s}(T_m)\), the agent \textbf{simultaneously} takes two actions:
\begin{itemize}[leftmargin=*]
    \item \textbf{Interval Selection} (\(\mathbf{a}_\ell(T_m)\)): a \textbf{discrete action} that chooses the rebalancing interval \(h_m \in \mathcal{H} = \{h_1, h_2, \dots, h_L\}\), where \(\mathcal{H}\) is a predefined set of candidate intervals (e.g., \{1, 5, 20\} days).
    \item \textbf{Portfolio Allocation} (\(\mathbf{a}_\rho(T_m)\)): a \textbf{continuous action} that determines the portfolio weight vector \(\mathbf{w}(T_m) \in \mathbb{R}^N\), subject to:
    \[
    w_i(T_m) \ge 0 \quad \text{and} \quad \sum_{i=1}^{N} w_i(T_m) = 1.
    \]
\end{itemize}

Once the agent executes action \((h_m, \mathbf{w}(T_m))\) at \(T_m\), the next decision point becomes:
\[
T_{m+1} = T_m + h_m.
\]

\paragraph{Price Relative Vector.}
To compute the portfolio value change over the interval \([T_m, T_{m+1})\), we define the \emph{price relative vector}:
\[
y(T_m) = \left[
\frac{p^{\text{close}}_{1, T_{m+1}}}{p^{\text{close}}_{1, T_m}},
\cdots,
\frac{p^{\text{close}}_{N, T_{m+1}}}{p^{\text{close}}_{N, T_m}}
\right],
\]
where \(p^{\text{close}}_{i,t}\) denotes the closing price of asset \(x_i\) at time \(t\).

\paragraph{Portfolio Return and Reward Signal.}
Given portfolio weights \(\mathbf{w}(T_m)\), we define the portfolio return over the rebalancing interval \(h_m\) as:
\[
R(T_m, h_m) = y(T_m) \cdot \mathbf{w}(T_m) - 1.
\]
While \(R(T_m, h_m)\) is not directly required to compute portfolio value updates, it serves as the core quantity for reinforcement learning: the agent's reward \(r_m\) is defined as a shaped version of this return, optionally adjusted with a bonus factor to promote optimal interval decisions:
\[
r_m = 
\begin{cases}
R(T_m, h_m) \cdot (1 + b), & \text{if } h_m = h^*(T_m), \\\\
R(T_m, h_m) \cdot (1 - b), & \text{otherwise},
\end{cases}
\]
where \(h^*(T_m)\) is the best-performing interval among candidates at time \(T_m\), computed \emph{ex-post} by evaluating the realized returns across all candidate intervals \(h \in \mathcal{H}\) based on observed market data and \(b\) is a fixed bonus coefficient (e.g., 0.2). This reward shaping encourages the agent to align its rebalancing frequency with market conditions, thereby improving long-term portfolio performance.

\paragraph{Transaction Costs.}
Rebalancing incurs transaction costs due to changes in portfolio weights. Let \(\mathbf{w}'(T_m)\) be the pre-rebalancing weights evolved from the previous allocation \(\mathbf{w}(T_{m-1})\) via:
\[
\mathbf{w}'(T_m) = \frac{y(T_{m-1}) \odot \mathbf{w}(T_{m-1})}{y(T_{m-1}) \cdot \mathbf{w}(T_{m-1})}.
\]
Then, the transaction cost factor \(\mu(T_m)\) is computed as:
\[
\mu(T_m) = 1 - c \sum_{i=1}^{N} |w_i(T_m) - w_i'(T_m)|,
\]
where \(c\) is the transaction cost rate (e.g., 0.002).

\paragraph{Portfolio Value Update.}
The actual portfolio value is updated using the price-relative return, as:
\[
V(T_{m+1}) = \mu(T_m) \cdot V(T_m) \cdot (y(T_m) \cdot \mathbf{w}(T_m)).
\]

\paragraph{Objective.}
The RL agent's goal is to maximize the final portfolio value \(V(T_M)\) at the last decision epoch \(T_M\). The overall optimization problem is:
\[
\max_{\{\mathbf{a}_\ell(T_m),\, \mathbf{a}_\rho(T_m)\}_{m=0}^{M}} V(T_M), \quad \text{subject to} \quad T_{m+1} = T_m + h_m.
\]
This formulation allows the policy to jointly optimize rebalancing frequency and allocation in a market-adaptive manner.

\section{Proposed Method}
\label{sec:method}

We propose \modelname, a deep reinforcement learning framework that jointly determines \emph{when to rebalance} (i.e., the trading interval) and \emph{how to allocate} (i.e., portfolio weights) in an end-to-end fashion. Unlike conventional approaches that assume fixed trading frequency, \modelname~adaptively selects the optimal rebalancing horizon based on evolving market conditions.

\subsection{Challenges in Portfolio Selection}
\label{subsec:challenges}

\begin{itemize}
    \item[\textbf{C1}] \textbf{Complexity of Stock Market Data.}  
    Stock market data is high-dimensional and non-stationary, with intricate cross-asset dependencies~\citep{jiang2017deep,lim2019enhancing,zhang2020cost}. Capturing both temporal patterns (e.g., momentum, seasonality) and cross-sectional relationships (e.g., correlations) is critical for effective portfolio selection.
    
    \item[\textbf{C2}] \textbf{Limitations of Fixed Rebalancing Intervals.}  
    Fixed rebalancing intervals (e.g., daily or weekly) may be suboptimal. In volatile markets, more frequent trades can help reduce drawdowns~\cite{Dichtl2014,Dai2024}, whereas in stable regimes, infrequent rebalancing can lower transaction costs~\citep{moody1998performance,jiang2017deep}.
    
    \item[\textbf{C3}] \textbf{Instability in Joint Action Learning.}  
    Learning both a discrete (rebalancing interval) and a continuous (portfolio allocation) action in a single policy can lead to unstable training. Robust RL algorithms are needed to jointly optimize these decisions.
\end{itemize}

\subsection{Key Ideas in \modelname}
\label{subsec:ideas}

\begin{itemize}
    \item[\textbf{Idea 1}] \textbf{Exploring Diverse Transformer Architectures (Addresses C1).}  
    We explore a range of transformer variants (e.g., the classical Transformer~\cite{vaswani2017attention}, Informer~\cite{zhou2021informerefficienttransformerlong}, \\
    Reformer~\cite{kitaev2020reformerefficienttransformer}, Autoformer~\cite{wu2021autoformer}) that effectively capture both temporal and cross-sectional dependencies via multi-head self-attention.
    
    \item[\textbf{Idea 2}] \textbf{Adaptive Interval Selection (Addresses C2).}  
    We introduce a discrete policy that adaptively selects the next rebalancing interval from a finite set \(\mathcal{H}=\{h_1,\dots,h_L\}\). For brevity, we denote the network implementing this function as \( f_{\mathrm{adapt}} \), which maps the asset embedding to the logits for horizon selection.
    
    \item[\textbf{Idea 3}] \textbf{Portfolio Allocation via PPO (Addresses C3).}  
    To jointly learn the rebalancing interval and portfolio weights, we employ PPO\cite{schulman2017ppo}. We denote the portfolio allocation network as \( f_{\mathrm{port}} \), which outputs parameters for a Gaussian policy over the portfolio weights.
\end{itemize}

Below, we describe how these ideas are implemented in \modelname, with notation consistent with Section~\ref{sec:problem}.

\subsection{Algorithm}
\label{subsec:algorithm}

Algorithm~\ref{alg:proposed_method} summarizes the training process. At each decision epoch \(t\), a rebalancing interval \(h \in \mathcal{H}\) is selected, and portfolio weights are determined based on the extracted features.

\begin{algorithm}[t]
\caption{Training Procedure for \modelname}
\label{alg:proposed_method}
\begin{algorithmic}[1]
\Require Historical data $\{\mathbf{X}(t)\}_{t=1}^{T} \in \mathbb{R}^{N \times \tau \times F}$; interval set $\mathcal{H}$; initial value $V_0$; Transformer $M_\theta$; PPO hyperparameters $(\gamma, \varepsilon, \beta, \alpha_v)$; bonus ratio $b$; max episodes $E$
\For{$\text{episode} = 1, \dots, E$}
    \State Initialize $V \gets V_0,\; t \gets 0$
    \While{$t < T$}
        \State \textbf{Feature Extraction:}
        \[
        \mathbf{e}(t) = \mathrm{FeedForwardBlock}\big(\mathrm{TemporalAttn}(M_\theta(\mathbf{X}(t)))\big)
        \]
        \State \textbf{Adaptive Interval Selection:}
        \[
        h \sim \mathrm{softmax}\left(f_{\mathrm{adapt}}(\mathbf{e}(t))\right)
        \]
        \State \textbf{Portfolio Allocation:}
        \[
        \tilde{\mathbf{a}}(t) \sim \mathcal{N}\left(\mu(t),\, \mathrm{diag}(\sigma(t)^2)\right)
        \]
        \[
        \mathbf{w}(t) = \mathrm{Normalize}\left(\tanh\left(\tilde{\mathbf{a}}(t)\right)\right)
        \]
        \State \textbf{Environment Transition:}
        \[
        R_t = y(t) \cdot \mathbf{w}(t) - 1,\quad V \gets \mu(t) \cdot V \cdot (1 + R_t)
        \]
        \State \textbf{Reward Adjustment:}
        \[
        r \gets 
        \begin{cases}
            R_t \cdot (1 + b), & \text{if } h = h^\ast \\
            R_t \cdot (1 - b), & \text{otherwise}
        \end{cases}
        \]
        \State Store transition $\left(\mathbf{X}(t), h, \mathbf{w}(t), r, \mathbf{X}(t + h)\right)$
        \State Update $t \gets t + h$
    \EndWhile
    \State \textbf{Policy Optimization:}
    \[
    \mathcal{L}_{\mathrm{total}} = \mathcal{L}_{\mathrm{PPO}} + \alpha_v \cdot \mathcal{L}_{\mathrm{value}}
    \]
\EndFor
\end{algorithmic}
\end{algorithm}

\subsection{Implementation Details}
\label{subsec:detailed_implementation}

Let \(\mathbf{X}(t) \in \mathbb{R}^{N \times \tau \times F}\) denote the historical data at decision epoch \(t\). We process \(\mathbf{X}(t)\) through the Transformer-based encoder \(M_\theta\) (which has a predictive interval \(T_{\mathrm{pred}}\) – i.e., it generates predictions or embeddings for the next \(T_{\mathrm{pred}}\) time steps – and a hidden size \(d_{\mathrm{model}}\)) to generate hidden representations:
\[
\mathbf{H}(t) \in \mathbb{R}^{N \times T_{\mathrm{pred}} \times d_{\mathrm{model}}}.
\]

A temporal attention mechanism is then applied:
\[
\widetilde{\mathbf{H}}(t) = \mathrm{TemporalAttn}\bigl(\mathbf{H}(t)\bigr),
\]
followed by a feed-forward block:
\begin{equation}
\label{eq:feed_forward_block}
\mathbf{e}(t) = \mathrm{FeedForwardBlock}\bigl(\widetilde{\mathbf{H}}(t)\bigr), \quad \mathbf{e}(t) \in \mathbb{R}^{N \times d_s}.
\end{equation}
This yields an embedding \(\mathbf{e}(t)\) for each asset, which is used for both adaptive interval selection and portfolio allocation.

\paragraph{Adaptive Interval Selection.}  
The function \(f_{\mathrm{adapt}}\) maps \(\mathbf{e}(t)\) to logits \(\mathbf{z}^{(\mathrm{adapt})}(t) \in \mathbb{R}^{L}\). After applying the softmax, we obtain a probability distribution \(p_t(\ell)\) over candidate intervals in \(\mathcal{H} = \{h_1, \dots, h_L\}\). A sample \(h\) drawn from \(p_t(\cdot)\) determines the next decision epoch \(t+h\).

\begin{table*}[ht]
\caption{
\textbf{Comparison of portfolio performance} across four markets (DJ 30, FTSE 100, KOSPI, CSI 300). 
Each row reports five metrics: 
\(\text{CAGR(\%)}\), \(\text{SR}\) (Sharpe Ratio), 
\(\text{SoR}\) (Sortino Ratio), 
\(\text{CR}\) (Calmar Ratio), 
and \(\text{MDD(\%)}\) (Maximum Drawdown). 
\textbf{Higher values} are better for \(\text{CAGR}\), \(\text{SR}\), \(\text{SoR}\), and \(\text{CR}\), whereas \textbf{lower values} are better for \(\text{MDD}\). 
Within each column, the \textbf{best} value is in \textbf{bold}, and the \underline{second-best} value is underlined. 
}
\label{table:main}
\centering
\resizebox{0.95\textwidth}{!}{%
\begin{tabular}{l rrrrr rrrrr rrrrr rrrrr}
\toprule
 & \multicolumn{5}{c}{\textbf{DJ 30}} 
 & \multicolumn{5}{c}{\textbf{FTSE 100}}
 & \multicolumn{5}{c}{\textbf{KOSPI}}
 & \multicolumn{5}{c}{\textbf{CSI 300}} \\

& \textbf{CAGR(\%)$\uparrow$} & \textbf{SR}$\uparrow$ & \textbf{SoR}$\uparrow$ & \textbf{CR}$\uparrow$ & \textbf{MDD(\%)}$\downarrow$
      & \textbf{CAGR(\%)}$\uparrow$ & \textbf{SR}$\uparrow$ & \textbf{SoR}$\uparrow$ & \textbf{CR}$\uparrow$ & \textbf{MDD(\%)}$\downarrow$
      & \textbf{CAGR(\%)}$\uparrow$ & \textbf{SR}$\uparrow$ & \textbf{SoR}$\uparrow$ & \textbf{CR}$\uparrow$ & \textbf{MDD(\%)}$\downarrow$
      & \textbf{CAGR(\%)}$\uparrow$ & \textbf{SR}$\uparrow$ & \textbf{SoR}$\uparrow$ & \textbf{CR}$\uparrow$ & \textbf{MDD(\%)}$\downarrow$
      \\
\midrule

\textbf{Index} 
 & 6.15 & 0.025 & 0.037 & 0.280 & 21.94
 & 4.27 & 0.019 & 0.025 & 0.387 & \textbf{11.03}
 & -5.33 & -0.019 & -0.029 & -0.153 & 34.79
 & -13.66 & -0.053 & -0.080 & -0.391 & 34.96 \\

\textbf{CSM} 
 & -18.58 & -0.113 & -0.157 & -0.373 & 49.86
 & -17.14 & -0.072 & -0.094 & -0.364 & 47.04
 & -49.56 & -0.117 & -0.176 & -0.624 & 79.41
 & -32.65 & -0.073 & -0.103 & -0.514 & 63.49 \\

\textbf{BLSW} 
 & -18.93 & -0.095 & -0.150 & -0.366 & 51.72
 & -19.13 & -0.062 & -0.103 & -0.414 & 46.17
 & -6.29 & -0.009 & -0.015 & -0.130 & 48.43
 & -18.97 & -0.051 & -0.081 & -0.404 & 46.93 \\

\midrule

\textbf{EIIE~\cite{jiang2017deep}} 
 & 0.11 & -0.002 & -0.004 & 0.005 & 22.90
 & \underline{9.86} & \underline{0.050} & \underline{0.069} & \textbf{0.658} & 14.98
 & 0.38 & 0.002 & 0.002 & 0.017 & \textbf{21.92}
 & -5.00 & -0.029 & -0.043 & -0.306 & \underline{16.32} \\

\textbf{DeepTrader~\cite{wang2021deeptrader}} 
 & \textbf{8.88} & \underline{0.035} & \underline{0.052} & \underline{0.572} & \underline{15.53}
 & 9.58 & 0.046 & \underline{0.069} & \underline{0.606} & 15.83
 & 7.06 & 0.023 & 0.029 & 0.215 & 32.84
 & -8.69 & -0.039 & -0.057 & -0.364 & 23.88 \\
 
\textbf{DeepClair~\cite{Choi_2024}} 
 & 4.35 & 0.018 & 0.027 & 0.191 & 16.17
 & 5.28 & 0.028 & 0.044 & 0.445 & \underline{11.85}
 & \underline{9.63} & \underline{0.029} & \underline{0.037} & \underline{0.303} & 31.75
 & \underline{-0.88} & \underline{-0.001} & \underline{-0.001} & \underline{-0.0043} & 20.69 \\

\midrule

\textbf{\modelname} 
 & \underline{7.90} & \textbf{0.130} & \textbf{0.211} & \textbf{0.577} & \textbf{13.67}
 & \textbf{9.95} & \textbf{0.086} & \textbf{0.123} & 0.546 & 18.23
 & \textbf{14.27} & \textbf{0.180} & \textbf{0.372} & \textbf{0.627} & \underline{22.77}
 & \textbf{5.09} & \textbf{0.070} & \textbf{0.109} & \textbf{0.465} & \textbf{10.95}\\

\bottomrule
\end{tabular}
}

\end{table*}

\begin{table*}[htbp]
\centering
\caption{Experimental results based on comparison between different rebalancing interval selection methods. \textbf{Fixed Daily}, \textbf{Fixed Daily}, \textbf{Fixed Monthly} refers to fixing the rebalancing interval to 1, 5, 20 at each market timestamp ($h_{\cdot}=1,5,20$) respectively. \textbf{Adaptive (20)} refers to selecting the rebalancing interval from 1 to 20 ($h_{\cdot}\in\{1,2,\cdots,20\}$. Our method adaptively selects the rebalancing interval from a more discrete range ($h_{\cdot}\in\{1,5,20\}$). Best scores are bold-faced while second-best ones are underlined.
}
\label{table:sub_interval}
\resizebox{0.95\textwidth}{!}{%
  \begin{tabular}{l c c c c c c c c c c c c c c c c c c c c}
    \hline
    \multicolumn{1}{c}{\textbf{Rebalancing}} 
      & \multicolumn{5}{c}{\textbf{DJ 30}} 
      & \multicolumn{5}{c}{\textbf{FTSE 100}} 
      & \multicolumn{5}{c}{\textbf{KOSPI}} 
      & \multicolumn{5}{c}{\textbf{CSI 300}} \\
    \multicolumn{1}{c}{\textbf{Interval}}
     & \textbf{CAGR(\%)$\uparrow$} & \textbf{SR}$\uparrow$ & \textbf{SoR}$\uparrow$ & \textbf{CR}$\uparrow$ & \textbf{MDD(\%)}$\downarrow$
      & \textbf{CAGR(\%)}$\uparrow$ & \textbf{SR}$\uparrow$ & \textbf{SoR}$\uparrow$ & \textbf{CR}$\uparrow$ & \textbf{MDD(\%)}$\downarrow$
      & \textbf{CAGR(\%)}$\uparrow$ & \textbf{SR}$\uparrow$ & \textbf{SoR}$\uparrow$ & \textbf{CR}$\uparrow$ & \textbf{MDD(\%)}$\downarrow$
      & \textbf{CAGR(\%)}$\uparrow$ & \textbf{SR}$\uparrow$ & \textbf{SoR}$\uparrow$ & \textbf{CR}$\uparrow$ & \textbf{MDD(\%)}$\downarrow$ \\
    \hline
    \textbf{Fixed Daily}     
      & 2.31  & 0.048  & 0.101  & 0.106  & 21.81 
      & 2.51  & 0.051  & 0.099  & 0.114  & 22.07  
      & -4.47 & -0.029 & -0.059 & -0.105 & 42.45 
      & 1.33  & 0.027  & 0.044  & 0.065  & 20.36 \\
    \textbf{Fixed Weekly}       
      & 1.69  & 0.035  & 0.081  & 0.091  & 18.71
      & \underline{6.24}  & 0.056  & 0.091  & 0.236  & 26.41
      & -3.50 & -0.052 & -0.106 & -0.137 & 25.62  
      & -1.03 & -0.014 & -0.025 & -0.066 & 15.11 \\
    \textbf{Fixed Monthly}      
      & \underline{5.19} & \underline{0.111}  & \textbf{0.230}  & \underline{0.251}  & \underline{20.71}  
      & 5.29  & \textbf{0.101}  & \textbf{0.155}  & \underline{0.433}  & \textbf{12.23}  
      & 5.71  & \underline{0.119}  & \underline{0.254}  & 0.266  & \textbf{21.45} 
      & \underline{3.98} & \textbf{0.078} & \textbf{0.200} & \underline{0.342} & \underline{11.66}  \\
    \textbf{Adaptive (20)}   
      & 3.90   & 0.051  & 0.095  & 0.167  & 23.32  
      & -2.88  & -0.039 & -0.057 & -0.152 & 18.91 
      & \underline{10.38}  & 0.101  & 0.199  & \underline{0.342}  & 30.33 
      & -13.75 & -0.28  & -0.495 & -0.383 & 35.94  \\
    \midrule
    \textbf{\modelname} & \textbf{7.90}  & \textbf{0.130} & \underline{0.211}  & \textbf{0.577} & \textbf{13.67}  
      & \textbf{9.95}  & \underline{0.086}  & \underline{0.123}  & \textbf{0.546} & \underline{18.23}  
      & \textbf{14.27} & \textbf{0.180} & \textbf{0.372} & \textbf{0.627} & \underline{22.77} 
      & \textbf{5.09}  & \underline{0.070}  & \underline{0.109}  & \textbf{0.465} & \textbf{10.95} \\
    \hline
  \end{tabular}%
}

\end{table*}


\paragraph{Portfolio Allocation.}  
Similarly, \(f_{\mathrm{port}}\) produces parameters \(\mu(t)\) and \(\sigma(t)\) (each in \(\mathbb{R}^{N}\)) for a Gaussian policy over portfolio weights. We then sample:
\[
\tilde{\mathbf{a}}(t) \sim \mathcal{N}\!\bigl(\mu(t),\,\mathrm{diag}(\sigma(t)^2)\bigr),
\]
apply a \(\tanh\) activation:
\[
\mathbf{a}(t) = \tanh\bigl(\tilde{\mathbf{a}}(t)\bigr),
\]
and normalize to obtain valid portfolio weights:
\[
\mathbf{w}(t) = \mathrm{Normalize}\bigl(\mathbf{a}(t)\bigr).
\]

\paragraph{Value Function Estimation.}  
To evaluate the expected return for each candidate rebalancing interval, we maintain a separate value function estimator \( \nu_\ell(\cdot) \) for each \(h_\ell \in \mathcal{H}\). The overall value estimate at state \(\mathbf{X}(t)\) is computed as a weighted sum:
\[
V^\pi\bigl(\mathbf{X}(t)\bigr) = \sum_{\ell=1}^{L} p_t(\ell) \; \nu_\ell\!\bigl(\mathbf{e}(t)\bigr).
\]
This formulation ensures that the policy accounts for the differing risk/return trade-offs associated with various rebalancing intervals.

\subsection{Loss Formulation}
\label{subsec:loss_formulation}

The objective of \modelname\ is to learn a trading policy that determines both \emph{when} to rebalance (through adaptive interval selection) and \emph{how} to allocate assets (through portfolio weight optimization), in order to maximize cumulative returns while accounting for transaction costs.

We define a joint policy as follows:
\[
\pi_\theta(\ell, \mathbf{a}(t) \mid \mathbf{X}(t)) = p_t(\ell) \cdot \pi_\theta(\mathbf{a}(t) \mid \mathbf{X}(t), \ell),
\]
where \(p_t(\ell)\) represents the probability of selecting a rebalancing interval \(h_\ell \in \mathcal{H}\), and \(\pi_\theta(\mathbf{a}(t) \mid \mathbf{X}(t), \ell)\) denotes a Gaussian policy over portfolio weights conditioned on the selected interval.

\textbf{To optimize this joint policy, we adopt the Proximal Policy Optimization (PPO) framework}~\cite{schulman2017ppo}. Since the action space consists of both a discrete decision (interval selection) and a continuous decision (portfolio allocation), we extend PPO to handle this composite structure. In particular:
\begin{itemize}
    \item The \textbf{discrete policy} \(p_t(\ell)\) is trained to select rebalancing intervals that maximize long-term expected rewards.
    \item The \textbf{continuous policy} over portfolio weights is optimized to improve returns within the chosen interval.
\end{itemize}

To further guide the learning of interval selection, we apply a reward shaping mechanism. Specifically, a bonus or penalty is applied depending on whether the selected interval \(h\) matches the best-performing horizon \(h^*\) at time \(t\):
\begin{equation}
\label{eq:bonus_reward}
r_t = 
\begin{cases}
R_t \cdot (1 + b), & \text{if } h = h^* \\
R_t \cdot (1 - b), & \text{otherwise},
\end{cases}
\end{equation}
where \(R_t\) is the portfolio return over \([t, t+h)\), and \(b > 0\) is a predefined bonus coefficient.

The overall loss function is defined as:
\[
\mathcal{L}_{\mathrm{total}} = \mathcal{L}_{\mathrm{PPO}} + \alpha_v\, \mathcal{L}_{\mathrm{value}},
\]
where \(\mathcal{L}_{\mathrm{PPO}}\) denotes the clipped policy gradient loss, and \(\mathcal{L}_{\mathrm{value}}\) is the mean squared error between the predicted and target returns. The coefficient \(\alpha_v\) controls the trade-off between policy improvement and value function accuracy.

This formulation enables \modelname\ to jointly learn both rebalancing schedules and allocation strategies that are responsive to changing market dynamics and investment horizons.

\section{Experiments}
\label{sec:experiments}

\subsection{Experiment Settings}
\label{sec:exp_setup}

We extensively evaluate our method across four major markets: DJ 30 (U.S.), FTSE 100 (Europe), KOSPI (Korea), and CSI 300 (China) over 20 years. For all experiments, we ran 10 independent trials with different random seeds and report the test results corresponding to the run with the lowest validation loss. Detailed experimental settings, including dataset splits and evaluation metrics, are provided in our interactive demo.

\subsection{\modelname~Overall Outperforms Other Methods in Portfolio Selection}
\label{sec:exp_q1}

\begin{table*}[]
\caption{Experimental results based on comparison between different combinations of Transformer variants and rebalancing interval settings (\textbf{Fixed}, \textbf{Adaptive}). \textbf{Fixed} refers to fixing rebalancing interval to 1 at each market timestep ($h_m=1$) while \textbf{Adaptive} refers to adaptively selecting the interval from $\mathcal{H}=\{1,5,20\}$. $\Delta$ refers to performance gain from replacing fixed, daily rebalancing with adaptive rebalancing interval selection for each Transformer variant setting. Color-coded shading in the $\Delta$ rows indicates positive (\colorbox{ACMRed!20}{red}) and negative (\colorbox{ACMBlue!20}{blue}) gains. Best scores are bold-faced while second-best ones are underlined.}
\label{table:models_types_markets_modified_final}
\resizebox{0.85\textwidth}{!}{ 
\begin{tabular}{llrrrrrrrrrrrr}
\toprule
 \multicolumn{1}{c}{\textbf{Transformer}} & \multicolumn{1}{c}{\textbf{Rebalancing}} & \multicolumn{3}{c}{\textbf{DJ 30}} & \multicolumn{3}{c}{\textbf{FTSE 100}} & \multicolumn{3}{c}{\textbf{KOSPI}} & \multicolumn{3}{c}{\textbf{CSI 300}} \\ 
 \multicolumn{1}{c}{\textbf{Variant}} & \multicolumn{1}{c}{\textbf{Interval}} & \textbf{CAGR(\%)$\uparrow$} & \textbf{SR$\uparrow$} & \textbf{MDD(\%)$\downarrow$} & \textbf{CAGR(\%)$\uparrow$} & \textbf{SR$\uparrow$} & \textbf{MDD(\%)$\downarrow$} & \textbf{CAGR(\%)$\uparrow$} & \textbf{SR$\uparrow$} & \textbf{MDD(\%)$\downarrow$} & \textbf{CAGR(\%)$\uparrow$} & \textbf{SR$\uparrow$} & \textbf{MDD(\%)$\downarrow$} \\
\midrule
\multirow{3}{*}{\textbf{LSTM (1997)\cite{Hochreiter1997lstm}}} 
 & \textbf{Fixed} 
   & 2.26 & 0.048 & 19.97
   & 2.81 & 0.050 & 17.23 
   & -4.61 & -0.014 & 37.51 
   & -11.32 & -0.039 & 36.17 \\
 & \textbf{Adaptive} 
   & 3.40 & 0.076 & 17.19 
   & 3.33 & 0.022 & 22.13 
   & -1.61 & -0.003 & 35.61 
   & -5.28 & -0.028 & 31.69 \\
 & $\bm{\triangle}$
   & \cellcolor{ACMRed!20}{1.14} & \cellcolor{ACMRed!20}{0.028} & \cellcolor{ACMRed!20}{-2.78}
   & \cellcolor{ACMRed!20}{0.52} & \cellcolor{ACMBlue!20}{-0.028} & \cellcolor{ACMBlue!20}{4.90}
   & \cellcolor{ACMRed!20}{3.00} & \cellcolor{ACMRed!20}{0.011} & \cellcolor{ACMRed!20}{-1.90}
   & \cellcolor{ACMRed!20}{6.04} & \cellcolor{ACMRed!20}{0.011} & \cellcolor{ACMRed!20}{-4.48} \\
\midrule

\multirow{3}{*}{\textbf{TCN (2016)\cite{lea2016tcn}}} 
 & \textbf{Fixed} 
   & 2.92 & 0.048 & \underline{15.19}\ 
   & 0.60 & 0.012 & 23.96 
   & -11.31 & -0.025 & 46.66 
   & -8.54 & -0.026 & 31.27 \\
 & \textbf{Adaptive} 
   & 4.80 & 0.101 & 17.42 
   & 1.70 & 0.008 & 23.42 
   & -11.69 & -0.026 & 49.96 
   & -3.75 & -0.013 & 27.15 \\
 & $\bm{\triangle}$ 
   & \cellcolor{ACMRed!20}{1.88} & \cellcolor{ACMRed!20}{0.053} & \cellcolor{ACMBlue!20}{2.23}
   & \cellcolor{ACMRed!20}{1.10} & \cellcolor{ACMBlue!20}{-0.004} & \cellcolor{ACMRed!20}{-0.54}
   & \cellcolor{ACMBlue!20}{-0.38} & \cellcolor{ACMRed!20}{0.001} & \cellcolor{ACMBlue!20}{3.30}
   & \cellcolor{ACMRed!20}{4.79} & \cellcolor{ACMRed!20}{0.013} & \cellcolor{ACMRed!20}{-4.12} \\
\midrule

\multirow{3}{*}{\textbf{Transformer (2017)\cite{vaswani2017attention}}} 
 & \textbf{Fixed} 
   & 4.83 & 0.032 & 20.92 
   & 3.73 & 0.016 & 29.37 
   & 2.77 & 0.012 & 35.21 
   & 3.59 & 0.069 & 15.31 \\
 & \textbf{Adaptive} 
   & 6.42 & 0.091 & 22.14 
   & 7.32 & \textbf{0.174} & \textbf{8.12} 
   & \underline{11.79} & \underline{0.136} & 23.17 
   & \textbf{17.53} & \textbf{0.220} & 13.38 \\
 & $\bm{\triangle}$ 
   & \cellcolor{ACMRed!20}{1.59} & \cellcolor{ACMRed!20}{0.059} & \cellcolor{ACMBlue!20}{1.22}
   & \cellcolor{ACMRed!20}{3.59} & \cellcolor{ACMRed!20}{0.158} & \cellcolor{ACMRed!20}{-21.25}
   & \cellcolor{ACMRed!20}{9.02} & \cellcolor{ACMRed!20}{0.124} & \cellcolor{ACMRed!20}{-12.04}
   & \cellcolor{ACMRed!20}{13.94} & \cellcolor{ACMRed!20}{0.151} & \cellcolor{ACMRed!20}{-1.93} \\
\midrule

\multirow{3}{*}{\textbf{Reformer (2020)\cite{kitaev2020reformerefficienttransformer}}} 
 & \textbf{Fixed} 
   & 3.81 & 0.017 & 25.76 
   & 4.80 & 0.022 & 19.39 
   & 0.28 & 0.005 & 39.17 
   & -0.69 & -0.001 & 20.54 \\
 & \textbf{Adaptive} 
   & 5.31 & 0.091 & 20.02 
   & 3.66 & 0.072 & 22.20 
   & 11.69 & 0.123 & 27.52 
   & 4.15 & 0.060 & 16.78 \\
 & $\bm{\triangle}$
   & \cellcolor{ACMRed!20}{1.50} & \cellcolor{ACMRed!20}{0.074} & \cellcolor{ACMRed!20}{-5.74}
   & \cellcolor{ACMBlue!20}{-1.14} & \cellcolor{ACMRed!20}{0.050} & \cellcolor{ACMBlue!20}{2.81}
   & \cellcolor{ACMRed!20}{11.41} & \cellcolor{ACMRed!20}{0.118} & \cellcolor{ACMRed!20}{-11.65}
   & \cellcolor{ACMRed!20}{4.84} & \cellcolor{ACMRed!20}{0.061} & \cellcolor{ACMRed!20}{-3.76} \\
\midrule

\multirow{3}{*}{\textbf{Informer (2021)\cite{zhou2021informerefficienttransformerlong}}} 
 & \textbf{Fixed} 
   & 6.34 & \textbf{0.27} & 18.53 
   & 3.36 & 0.014 & 27.19 
   & -8.03 & -0.017 & 48.72 
   & 0.12 & 0.003 & 31.00 \\
 & \textbf{Adaptive} 
   & 3.50 & 0.074 & 22.60 
   & 1.56 & 0.032 & 13.50 
   & -5.94 & -0.040 & 46.43 
   & \underline{14.22} & 0.074 & 15.27 \\
 & $\bm{\triangle}$
   & \cellcolor{ACMBlue!20}{-2.84} & \cellcolor{ACMBlue!20}{-0.196} & \cellcolor{ACMBlue!20}{4.07}
   & \cellcolor{ACMBlue!20}{-1.80} & \cellcolor{ACMRed!20}{0.018} & \cellcolor{ACMRed!20}{-13.69}
   & \cellcolor{ACMRed!20}{2.09} & \cellcolor{ACMBlue!20}{-0.023} & \cellcolor{ACMRed!20}{-2.29}
   & \cellcolor{ACMRed!20}{14.10} & \cellcolor{ACMRed!20}{0.071} & \cellcolor{ACMRed!20}{-15.73} \\
\midrule

\multirow{3}{*}{\textbf{Autoformer (2021)\cite{wu2021autoformer}}} 
 & \textbf{Fixed} 
   & \textbf{7.93} & 0.034 & 21.50 
   & 1.59 & 0.008 & 25.33 
   & -9.87 & -0.033 & 44.65 
   & -0.66 & 0.002 & \underline{22.59} \\
 & \textbf{Adaptive} 
   & 3.86 & 0.083 & 20.20 
   & 6.10 & 0.117 & 16.19 
   & 5.62 & 0.047 & 32.16 
   & -1.12 & -0.001 & 25.75 \\                                                                                                                                                       
 & $\bm{\triangle}$ 
   & \cellcolor{ACMBlue!20}{-4.07} & \cellcolor{ACMRed!20}{0.049} & \cellcolor{ACMRed!20}{-1.30}
   & \cellcolor{ACMRed!20}{4.51} & \cellcolor{ACMRed!20}{0.109} & \cellcolor{ACMRed!20}{-9.14}
   & \cellcolor{ACMRed!20}{15.49} & \cellcolor{ACMRed!20}{0.080} & \cellcolor{ACMRed!20}{-12.49}
   & \cellcolor{ACMBlue!20}{-0.46} & \cellcolor{ACMBlue!20}{-0.003} & \cellcolor{ACMBlue!20}{3.16} \\
\midrule

\multirow{3}{*}{\textbf{Flashformer (2022)\cite{dao2022flashattentionfastmemoryefficientexact}}} 
 & \textbf{Fixed} 
   & 7.14 & 0.030 & 20.70 
   & \underline{8.09} & 0.031 & 20.64 
   & -6.55 & -0.013 & 51.39 
   & 5.34 & 0.020 & 22.99 \\
 & \textbf{Adaptive} 
   & 5.62 & 0.116 & 16.23 
   & 7.14 & \underline{0.129} & \underline{13.26} 
   & 11.79 & 0.045 & 30.58 
   & 12.44 & \underline{0.217} & \underline{11.21} \\
 & $\bm{\triangle}$ 
   & \cellcolor{ACMBlue!20}{-1.52} & \cellcolor{ACMRed!20}{0.086} & \cellcolor{ACMRed!20}{-4.47}
   & \cellcolor{ACMBlue!20}{-0.95} & \cellcolor{ACMRed!20}{0.098} & \cellcolor{ACMRed!20}{-7.38}
   & \cellcolor{ACMRed!20}{18.34} & \cellcolor{ACMRed!20}{0.058} & \cellcolor{ACMRed!20}{-20.81}
   & \cellcolor{ACMRed!20}{7.10} & \cellcolor{ACMRed!20}{0.197} & \cellcolor{ACMRed!20}{-11.78} \\
\midrule

\multirow{3}{*}{\textbf{FEDformer (2022)\cite{zhou2022fedformer}}} 
 & \textbf{Fixed} 
   & 5.07 & 0.049 & 20.31 
   & 2.92 & 0.013 & 26.80 
   & 0.82 & 0.013 & \underline{23.09} 
   & 2.99 & 0.013 & 24.35 \\
 & \textbf{Adaptive} 
   & 4.16 & 0.050 & 22.82 
   & 4.91 & 0.085 & 15.57 
   & 3.19 & 0.060 & 37.51 
   & 5.41 & 0.100 & 14.18 \\
 & $\bm{\triangle}$ 
   & \cellcolor{ACMBlue!20}{-0.91} & \cellcolor{ACMRed!20}{0.001} & \cellcolor{ACMBlue!20}{2.51}
   & \cellcolor{ACMRed!20}{1.99} & \cellcolor{ACMRed!20}{0.072} & \cellcolor{ACMRed!20}{-11.23}
   & \cellcolor{ACMRed!20}{2.37} & \cellcolor{ACMRed!20}{0.047} & \cellcolor{ACMBlue!20}{14.42}
   & \cellcolor{ACMRed!20}{2.42} & \cellcolor{ACMRed!20}{0.087} & \cellcolor{ACMRed!20}{-10.17} \\
\midrule

\multirow{3}{*}{\textbf{Crossformer (2023)\cite{zhang2023crossformer}}} 
 & \textbf{Fixed} 
   & 5.49 & 0.023 & 23.74 
   & 3.61 & 0.015 & 29.02 
   & 0.58 & 0.005 & 27.90 
   & -7.86 & -0.023 & 32.06 \\
 & \textbf{Adaptive} 
   & 5.03 & 0.103 & 18.82 
   & 6.19 & 0.060 & 20.41 
   & 9.32 & 0.069 & 27.02 
   & 9.48 & 0.034 & 20.79 \\
 & $\bm{\triangle}$ 
   & \cellcolor{ACMBlue!20}{-0.46} & \cellcolor{ACMRed!20}{0.080} & \cellcolor{ACMRed!20}{-4.92}
   & \cellcolor{ACMRed!20}{2.58} & \cellcolor{ACMRed!20}{0.045} & \cellcolor{ACMRed!20}{-8.61}
   & \cellcolor{ACMRed!20}{8.74} & \cellcolor{ACMRed!20}{0.064} & \cellcolor{ACMRed!20}{-0.88}
   & \cellcolor{ACMRed!20}{17.34} & \cellcolor{ACMRed!20}{0.057} & \cellcolor{ACMRed!20}{-11.27} \\
\midrule

\multirow{3}{*}{\shortstack{\textbf{iTransformer (2023)\cite{liu2023itransformer}} \\ \textbf{ (Ours)}}}
 & \textbf{Fixed} 
   & 2.31 & 0.048 & 21.81 
   & 2.51 & 0.051 & 22.07 
   & -4.47 & -0.029 & 42.45 
   & 1.33 & 0.027 & 20.36 \\
& \textbf{Adaptive} 
   & \underline{7.90} & \underline{0.130} & \textbf{13.67}
   & \textbf{9.95} & 0.086 & 18.23
   & \textbf{14.27} & \textbf{0.180} & \textbf{22.77}
   & 5.09 & 0.070 & \textbf{10.95} \\
 & $\Delta$ 
   & \cellcolor{ACMRed!20}{5.59} & \cellcolor{ACMRed!20}{0.082} & \cellcolor{ACMRed!20}{-8.14}
   & \cellcolor{ACMRed!20}{7.44} & \cellcolor{ACMRed!20}{0.035} & \cellcolor{ACMRed!20}{-3.84}
   & \cellcolor{ACMRed!20}{18.74} & \cellcolor{ACMRed!20}{0.209} & \cellcolor{ACMRed!20}{-19.68}
   & \cellcolor{ACMRed!20}{3.76} & \cellcolor{ACMRed!20}{0.043} & \cellcolor{ACMRed!20}{-9.41} \\
\bottomrule
\end{tabular}
}

\end{table*}

We first conducted experiments to examine whether our newly proposed \modelname~exhibits robustness in portfolio selection under various stock market environments compared to baseline methods. Specifically, we quantitatively assessed the benefits of integrating our key ideas—careful selection of the Transformer variant, adaptive rebalancing interval selection, and proximal policy optimization—into the implementation of \modelname.

As shown in Table~\ref{table:main}, our approach demonstrated outstanding performance by outperforming 17 out of 20 evaluation metrics across four different regional markets. In particular, \modelname~ consistently achieved the best results in \textbf{CAGR} (7.90\%, 9.95\%, 14.27\%, 5.09\%), \textbf{SR} (0.130, 0.086, 0.180, 0.109), and \textbf{SoR} (0.211, 0.123, 0.372, 0.109) in all markets. Indeed, these performance gains persist even under diverse market conditions (e.g., an upward-trending DJ 30 and a volatile KOSPI). Notably, while the baseline methods exhibited negative investment performance in the CSI 300 market—a market characterized by irregular stock price patterns that undermine the generalizability of conventional models—only \modelname~ excelled in generating positive risk-adjusted returns, as evidenced by its \textbf{CAGR} (5.09\%), \textbf{SR} (0.070), \textbf{SoR} (0.109), and \textbf{CR} (0.465).

These results conclusively highlight the potential benefits of adaptively selecting rebalancing intervals for each portfolio selection timestep. Whereas prior methods fix the interval to either daily or weekly basis, \modelname~dynamically adjusts its rebalancing frequency in response to market signals—likely contributing to the observed improvement in both return and drawdown metrics. To further investigate the contribution of this adaptive mechanism, we performed additional experiments focusing on specific model design components central to our approach.

\subsection{Adaptive Rebalancing Interval Selection Generates Better Investment Outcomes than Its Fixed Counterpart}
Our core approach in \modelname~is selecting rebalancing interval based on its percieved market state. We conducted ablation experiments by replacing the adaptive rebalancing interval selection component with fixed rebalancing interval method. That is, the ablated versions of \modelname~only select a fixed value of $h_{\cdot}$ at each market timestep. While fixed daily rebalancing (\textbf{Fixed Daily}, $h_{\cdot}=1$) is most commonly used in previously introduced methods, we also included weekly (\textbf{Fixed Weekly}, $h_{\cdot}=5$) and monthly (\textbf{Fixed Monthly}, $h_{\cdot}=20$) rebalancing strategies as both of them are plausible intervals in actual investment domains.

According to Table~\ref{table:sub_interval}, our adaptive rebalancing interval selection method outperformed \textbf{Fixed Daily} in all evaluation metrics across all market environments. This clearly shows that fixed daily rebalancing strategies have several limitations in maximizing risk-adjusted returns given different market circumstances, which is also evidenced in the results in Table~\ref{table:main}. Overall, our model achieved best results in 13 out of 25 evaluation metrics while the rest of them being second-best. Interestingly, \textbf{Fixed Monthly} demonstrated its robustness comparable with our model, suggesting that certain market dynamics may favor consistent long-term holding strategies. This is exemplified by its achieving the best \textbf{SoR} in three markets (0.230, 0.155, 0.078).

In conclusion, our findings demonstrate that adaptive rebalancing interval selection can significantly enhance portfolio performance by dynamically adjusting to market conditions. Moreover, the strong performance of fixed monthly rebalancing suggests that longer rebalancing intervals may offer a more resilient baseline than daily rebalancing, potentially mitigating the risks associated with short-term portfolio adjustments. 

\begin{figure*}
    \centering
    \includegraphics[width=0.75\textwidth]{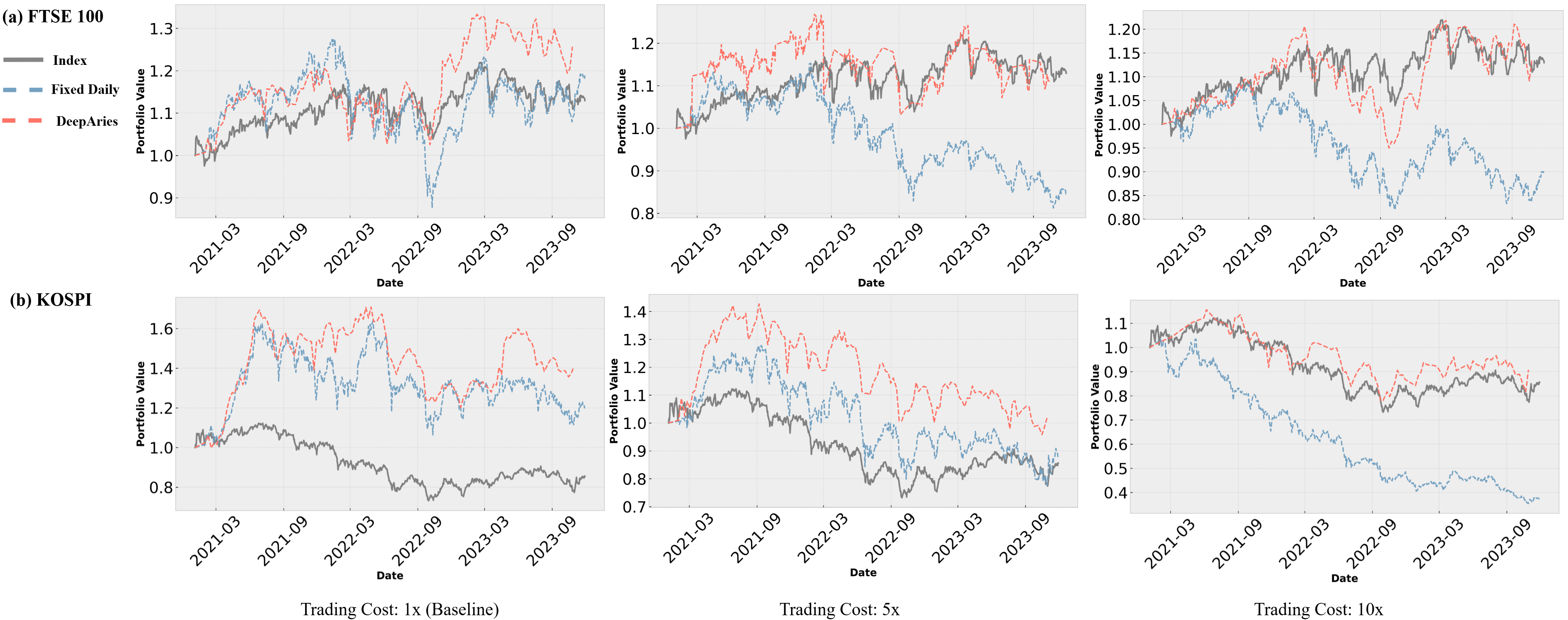}
    \caption{Effect of transaction costs on portfolio performance. The six panels are arranged in three columns and two rows. In each column, the transaction cost is fixed: the left column shows results with the baseline fee of 0.01\%, the middle column uses fees increased 5-fold, and the right column uses fees increased 10-fold. The top row displays results for the FTSE 100 (reflecting bullish market conditions), while the bottom row shows results for the KOSPI (reflecting volatile market conditions).}
    \label{fig:trading_cost}
\end{figure*}

\subsection{\modelname~with iTransformer Achieves Superior Performance via Adaptive Rebalancing}

Optimal interval selection demands effective exploitation of both temporal patterns and cross-sectional relationships in high dimensional stock market data. While earlier models have employed sequential modules such as LSTMs, TCNs, and Transformers, the latter have shown exceptional feature extraction capabilities that can enhance portfolio performance.

Given the rapid evolution of Transformer variants in forecasting tasks, selecting an appropriate backbone is crucial to avoid suboptimal reinforcement learning trajectories in our adaptive rebalancing framework. Prior portfolio selection models have largely relied on a limited range of Transformer architectures, potentially overlooking the benefits of more recent advances. This necessitates a comprehensive exploration through extensive ablation studies.

Inspired by DeepClair—which investigated four forecasting Transformers—we adopted a broad set of Transformer architectures (\textbf{Transformer}, \textbf{Reformer}, \textbf{Informer}, \textbf{Autoformer}, \textbf{Flashformer}, \textbf{FEDformer}, \textbf{Crossformer}, \textbf{iTransformer}) and conducted experiments across four major markets. To contrast the impact of adaptive rebalancing interval selection, we also ran experiments on the same variants with a fixed daily rebalancing interval (\textbf{Fixed}). Additionally, we included \textbf{LSTM} and \textbf{TCN} models, as they have been previously utilized in RL-based portfolio selection~\cite{jiang2017deep,borovykh2017temporal}.

Our investigation reveals that \textbf{iTransformer}, when paired with adaptive rebalancing, outperforms its fixed-interval counterpart. For instance, in the DJ 30 market, \textbf{iTransformer} achieves a CAGR improvement from 2.31\% (Fixed) to 7.90\% (Adaptive), with the Sharpe Ratio increasing from 0.048 to 0.130 and the Maximum Drawdown decreasing from 21.81\% to 13.67\%. Similar trends are observed across the FTSE 100, KOSPI, and CSI 300 markets. These results indicate that the adaptive mechanism enables \textbf{iTransformer} to more effectively capture market dynamics and adjust its portfolio strategy, thereby enhancing risk-adjusted returns and mitigating downside risks.

\subsection{Adaptive Rebalancing Selection Shows Enhanced Resilience Against Elevated Transaction Costs}

Transaction costs represent a critical factor in portfolio management, as they directly impact net returns. To evaluate the robustness of our adaptive rebalancing strategy against increased transaction costs, we conducted experiments comparing our model (\modelname) with a Fixed Daily rebalancing baseline. For this analysis, we focused on two representative markets: FTSE 100, which exhibits bullish tendencies, and KOSPI, characterized by heightened volatility.
In our experimental setup, the baseline transaction cost was set at 0.01\%, and we subsequently examined fee scenarios with costs increased by 5-fold and 10-fold. As illustrated in Figure~\ref{fig:trading_cost}, while higher transaction costs deteriorate the overall portfolio performance for both approaches, the Fixed Daily strategy is more adversely affected. In contrast, \modelname's adaptive rebalancing selection mitigates the negative impact of increased fees, resulting in relatively higher portfolio returns compared to the Fixed Daily baseline. This indicates that our adaptive strategy is more resilient to transaction cost variations—a critical advantage in real-world scenarios and interactive demos where dynamic asset adjustments and prolonged trading periods can amplify fee-related losses.


\section{Conclusion}
\label{sec:conclusion}

In this work, we introduced \modelname, a reinforcement learning framework that dynamically adjusts rebalancing intervals based on market conditions. Our approach departs from conventional fixed-interval strategies by integrating adaptive interval selection with Transformer-based forecasting to capture market dynamics and minimize transaction costs effectively. Extensive evaluations across major markets demonstrated that adaptively selecting rebalancing intervals significantly improves risk-adjusted returns and overall portfolio performance, and our interactive demo further highlights the practical utility of the proposed framework.




\begin{acks}
This work was supported by Institute of Information \& communications Technology Planning \& Evaluation (IITP) under the ICT Creative Consilience program (IITP-2025-RS-2020-II201819) and the Artificial Intelligence Convergence Innovation Human Resources Development (IITP-2025-RS-2023-00254177) grant funded by the Korea government(MSIT). This research was also supported by the National Research Foundation of Korea (NRF) [NRF2023R1A2C3004176].

Donghee Choi was supported by the CoDiet project at Imperial College London. The CoDiet project is funded by the European Union under Horizon Europe grant number 101084642. CoDiet research activities taking place at Imperial College London are supported by UK Research and Innovation (UKRI) under the UK government's Horizon Europe funding guarantee (grant number 101084642).

\end{acks}

\section{GenAI Usage Disclosure}

Generative AI tools were not used in the conception of the research or in the experimental process. 
They were employed only for minor editorial purposes, such as grammar checking and improving readability.



\bibliographystyle{ACM-Reference-Format}
\bibliography{ref}


\begin{thebibliography}{34}


\ifx \showCODEN    \undefined \def \showCODEN     #1{\unskip}     \fi
\ifx \showISBNx    \undefined \def \showISBNx     #1{\unskip}     \fi
\ifx \showISBNxiii \undefined \def \showISBNxiii  #1{\unskip}     \fi
\ifx \showISSN     \undefined \def \showISSN      #1{\unskip}     \fi
\ifx \showLCCN     \undefined \def \showLCCN      #1{\unskip}     \fi
\ifx \shownote     \undefined \def \shownote      #1{#1}          \fi
\ifx \showarticletitle \undefined \def \showarticletitle #1{#1}   \fi
\ifx \showURL      \undefined \def \showURL       {\relax}        \fi
\providecommand\bibfield[2]{#2}
\providecommand\bibinfo[2]{#2}
\providecommand\natexlab[1]{#1}
\providecommand\showeprint[2][]{arXiv:#2}

\bibitem[Almahdi and Yang(2017)]%
        {Almahdi2017}
\bibfield{author}{\bibinfo{person}{Saud Almahdi} {and} \bibinfo{person}{S.~Y. Yang}.} \bibinfo{year}{2017}\natexlab{}.
\newblock \showarticletitle{An Adaptive Portfolio Trading System: A Risk-Return Portfolio Optimization Using Recurrent Reinforcement Learning with Expected Maximum Drawdown}.
\newblock \bibinfo{journal}{\emph{Expert Systems with Applications}}  \bibinfo{volume}{87} (\bibinfo{year}{2017}), \bibinfo{pages}{267--279}.
\newblock


\bibitem[Benhamou(2023)]%
        {benhamou2023}
\bibfield{author}{\bibinfo{person}{Eric Benhamou}.} \bibinfo{year}{2023}\natexlab{}.
\newblock \bibinfo{title}{Can Deep Reinforcement Learning Solve the Portfolio Allocation Problem?}
\newblock
\urldef\tempurl%
\url{https://ssrn.com/abstract=4599800}
\showURL{%
\tempurl}
\newblock
\shownote{PhD Thesis Manuscript, SSRN 4599800}.


\bibitem[Borovykh et~al\mbox{.}(2017)]%
        {borovykh2017temporal}
\bibfield{author}{\bibinfo{person}{Anastasia Borovykh}, \bibinfo{person}{Sander Bohte}, {and} \bibinfo{person}{Cornelis~W. Oosterlee}.} \bibinfo{year}{2017}\natexlab{}.
\newblock \showarticletitle{Conditional Time Series Forecasting with Convolutional Neural Networks}.
\newblock \bibinfo{journal}{\emph{International Journal of Forecasting}} \bibinfo{volume}{33}, \bibinfo{number}{3} (\bibinfo{year}{2017}), \bibinfo{pages}{811--819}.
\newblock


\bibitem[Choi et~al\mbox{.}(2024)]%
        {Choi_2024}
\bibfield{author}{\bibinfo{person}{Donghee Choi}, \bibinfo{person}{Jinkyu Kim}, \bibinfo{person}{Mogan Gim}, \bibinfo{person}{Jinho Lee}, {and} \bibinfo{person}{Jaewoo Kang}.} \bibinfo{year}{2024}\natexlab{}.
\newblock \showarticletitle{DeepClair: Utilizing Market Forecasts for Effective Portfolio Selection}. In \bibinfo{booktitle}{\emph{Proceedings of the 33rd ACM International Conference on Information and Knowledge Management}}. \bibinfo{pages}{4414--4422}.
\newblock


\bibitem[Dai et~al\mbox{.}(2024)]%
        {Dai2024}
\bibfield{author}{\bibinfo{person}{Min Dai}, \bibinfo{person}{Fangyuan Yi}, {and} \bibinfo{person}{Yue Zhong}.} \bibinfo{year}{2024}\natexlab{}.
\newblock \showarticletitle{Optimal Dynamic Portfolio Rebalancing}.
\newblock \bibinfo{journal}{\emph{Journal of Economic Dynamics and Control}}  \bibinfo{volume}{158} (\bibinfo{year}{2024}), \bibinfo{pages}{104773}.
\newblock


\bibitem[Dao et~al\mbox{.}(2022)]%
        {dao2022flashattentionfastmemoryefficientexact}
\bibfield{author}{\bibinfo{person}{Tri Dao}, \bibinfo{person}{Daniel~Y. Fu}, \bibinfo{person}{Stefano Ermon}, \bibinfo{person}{Atri Rudra}, {and} \bibinfo{person}{Christopher Ré}.} \bibinfo{year}{2022}\natexlab{}.
\newblock \bibinfo{title}{FlashAttention: Fast and Memory-Efficient Exact Attention with IO-Awareness}.
\newblock
\showeprint[arxiv]{2205.14135}~[cs.LG]
\urldef\tempurl%
\url{https://arxiv.org/abs/2205.14135}
\showURL{%
\tempurl}


\bibitem[Daryanani(2008)]%
        {daryanani2008opportunistic}
\bibfield{author}{\bibinfo{person}{Gobind Daryanani}.} \bibinfo{year}{2008}\natexlab{}.
\newblock \showarticletitle{Opportunistic Rebalancing: A New Paradigm for Wealth Managers}.
\newblock \bibinfo{journal}{\emph{Journal of Financial Planning}} \bibinfo{volume}{21}, \bibinfo{number}{2} (\bibinfo{year}{2008}), \bibinfo{pages}{48--61}.
\newblock


\bibitem[Dichtl et~al\mbox{.}(2014)]%
        {Dichtl2014}
\bibfield{author}{\bibinfo{person}{Hubert Dichtl}, \bibinfo{person}{Wolfgang Drobetz}, {and} \bibinfo{person}{Manuel Wambach}.} \bibinfo{year}{2014}\natexlab{}.
\newblock \showarticletitle{Where Is the Value Added of Rebalancing? A Systematic Comparison of Alternative Rebalancing Strategies}.
\newblock \bibinfo{journal}{\emph{Financial Markets and Portfolio Management}} \bibinfo{volume}{28}, \bibinfo{number}{3} (\bibinfo{year}{2014}), \bibinfo{pages}{209--231}.
\newblock


\bibitem[Hochreiter and Schmidhuber(1997)]%
        {Hochreiter1997lstm}
\bibfield{author}{\bibinfo{person}{Sepp Hochreiter} {and} \bibinfo{person}{Jürgen Schmidhuber}.} \bibinfo{year}{1997}\natexlab{}.
\newblock \showarticletitle{Long Short-Term Memory}. In \bibinfo{booktitle}{\emph{Neural Computation}}, Vol.~\bibinfo{volume}{9}. \bibinfo{pages}{1735--1780}.
\newblock


\bibitem[Hwang et~al\mbox{.}(2023)]%
        {HWANG2023104285}
\bibfield{author}{\bibinfo{person}{Yoontae Hwang}, \bibinfo{person}{Junpyo Park}, \bibinfo{person}{Yongjae Lee}, {and} \bibinfo{person}{Dong-Young Lim}.} \bibinfo{year}{2023}\natexlab{}.
\newblock \showarticletitle{Stop-loss adjusted labels for machine learning-based trading of risky assets}.
\newblock \bibinfo{journal}{\emph{Finance Research Letters}}  \bibinfo{volume}{58} (\bibinfo{year}{2023}), \bibinfo{pages}{104285}.
\newblock
\showISSN{1544-6123}
\href{https://doi.org/10.1016/j.frl.2023.104285}{doi:\nolinkurl{10.1016/j.frl.2023.104285}}


\bibitem[Jiang et~al\mbox{.}(2017)]%
        {jiang2017deep}
\bibfield{author}{\bibinfo{person}{Zhengyao Jiang}, \bibinfo{person}{Dixing Xu}, {and} \bibinfo{person}{Jinjun Liang}.} \bibinfo{year}{2017}\natexlab{}.
\newblock \showarticletitle{A Deep Reinforcement Learning Framework for the Financial Portfolio Management Problem}.
\newblock \bibinfo{journal}{\emph{arXiv preprint arXiv:1706.10059}} (\bibinfo{year}{2017}).
\newblock


\bibitem[Kim et~al\mbox{.}(2023)]%
        {kim2023hadaps}
\bibfield{author}{\bibinfo{person}{Jinkyu Kim}, \bibinfo{person}{Donghee Choi}, \bibinfo{person}{Mogan Gim}, {and} \bibinfo{person}{Jaewoo Kang}.} \bibinfo{year}{2023}\natexlab{}.
\newblock \showarticletitle{HADAPS: Hierarchical Adaptive Multi-Asset Portfolio Selection}.
\newblock \bibinfo{journal}{\emph{IEEE Access}} (\bibinfo{year}{2023}).
\newblock


\bibitem[Kitaev et~al\mbox{.}(2020)]%
        {kitaev2020reformerefficienttransformer}
\bibfield{author}{\bibinfo{person}{Nikita Kitaev}, \bibinfo{person}{Łukasz Kaiser}, {and} \bibinfo{person}{Anselm Levskaya}.} \bibinfo{year}{2020}\natexlab{}.
\newblock \bibinfo{title}{Reformer: The Efficient Transformer}.
\newblock
\showeprint[arxiv]{2001.04451}~[cs.LG]
\urldef\tempurl%
\url{https://arxiv.org/abs/2001.04451}
\showURL{%
\tempurl}


\bibitem[Lea et~al\mbox{.}(2016)]%
        {lea2016tcn}
\bibfield{author}{\bibinfo{person}{Colin Lea}, \bibinfo{person}{Rene Vidal}, \bibinfo{person}{Austin Reiter}, {and} \bibinfo{person}{Gregory~D Hager}.} \bibinfo{year}{2016}\natexlab{}.
\newblock \showarticletitle{Temporal convolutional networks: A unified approach to action segmentation}. In \bibinfo{booktitle}{\emph{Computer Vision--ECCV 2016 Workshops: Amsterdam, The Netherlands, October 8-10 and 15-16, 2016, Proceedings, Part III}}. Springer, \bibinfo{pages}{47--54}.
\newblock


\bibitem[Li et~al\mbox{.}(2023)]%
        {Li2023}
\bibfield{author}{\bibinfo{person}{Xinyi Li}, \bibinfo{person}{Zhi Wang}, \bibinfo{person}{Jun Gao}, {and} \bibinfo{person}{Haifeng Zhu}.} \bibinfo{year}{2023}\natexlab{}.
\newblock \showarticletitle{Transformer-based Deep Reinforcement Learning for High-frequency Portfolio Management}.
\newblock \bibinfo{journal}{\emph{Expert Systems with Applications}}  \bibinfo{volume}{234} (\bibinfo{year}{2023}), \bibinfo{pages}{120778}.
\newblock


\bibitem[Lim et~al\mbox{.}(2020)]%
        {lim2019enhancing}
\bibfield{author}{\bibinfo{person}{Bryan Lim}, \bibinfo{person}{Stefan Zohren}, {and} \bibinfo{person}{Stephen Roberts}.} \bibinfo{year}{2020}\natexlab{}.
\newblock \bibinfo{title}{Enhancing Time Series Momentum Strategies Using Deep Neural Networks}.
\newblock
\showeprint[arxiv]{1904.04912}~[stat.ML]
\urldef\tempurl%
\url{https://arxiv.org/abs/1904.04912}
\showURL{%
\tempurl}


\bibitem[Liu et~al\mbox{.}(2020)]%
        {Liu2020}
\bibfield{author}{\bibinfo{person}{Xiao-Yang Liu}, \bibinfo{person}{Hongyang Yang}, \bibinfo{person}{Qian Chen}, \bibinfo{person}{Runjia Zhang}, \bibinfo{person}{Liuqing Wu}, {and} \bibinfo{person}{Bowen Wang}.} \bibinfo{year}{2020}\natexlab{}.
\newblock \showarticletitle{FinRL: A Deep Reinforcement Learning Library for Automated Stock Trading in Quantitative Finance}.
\newblock \bibinfo{journal}{\emph{arXiv preprint arXiv:2011.09607}} (\bibinfo{year}{2020}).
\newblock


\bibitem[Liu et~al\mbox{.}(2023)]%
        {liu2023itransformer}
\bibfield{author}{\bibinfo{person}{Yong Liu}, \bibinfo{person}{Tengge Hu}, \bibinfo{person}{Haoran Zhang}, \bibinfo{person}{Haixu Wu}, \bibinfo{person}{Shiyu Wang}, \bibinfo{person}{Lintao Ma}, {and} \bibinfo{person}{Mingsheng Long}.} \bibinfo{year}{2023}\natexlab{}.
\newblock \showarticletitle{iTransformer: Inverted Transformers Are Effective for Time Series Forecasting}.
\newblock \bibinfo{journal}{\emph{arXiv preprint arXiv:2310.06625}} (\bibinfo{year}{2023}).
\newblock


\bibitem[Markowitz(1952)]%
        {markowitz1952portfolio}
\bibfield{author}{\bibinfo{person}{Harry Markowitz}.} \bibinfo{year}{1952}\natexlab{}.
\newblock \showarticletitle{Portfolio Selection}.
\newblock \bibinfo{journal}{\emph{The Journal of Finance}} \bibinfo{volume}{7}, \bibinfo{number}{1} (\bibinfo{year}{1952}), \bibinfo{pages}{77--91}.
\newblock


\bibitem[Moody and Wu(1998)]%
        {moody1998performance}
\bibfield{author}{\bibinfo{person}{John Moody} {and} \bibinfo{person}{L. Wu}.} \bibinfo{year}{1998}\natexlab{}.
\newblock \showarticletitle{Performance Functions and Reinforcement Learning for Trading Systems and Portfolios}.
\newblock \bibinfo{journal}{\emph{Journal of Forecasting}} \bibinfo{volume}{17}, \bibinfo{number}{5-6} (\bibinfo{year}{1998}), \bibinfo{pages}{441--470}.
\newblock


\bibitem[Olivares-Nadal and DeMiguel(2018)]%
        {OlivaresNadal2018}
\bibfield{author}{\bibinfo{person}{Alba~V. Olivares-Nadal} {and} \bibinfo{person}{Victor DeMiguel}.} \bibinfo{year}{2018}\natexlab{}.
\newblock \showarticletitle{A Robust Perspective on Transaction Costs in Portfolio Optimization}.
\newblock \bibinfo{journal}{\emph{Operations Research}} \bibinfo{volume}{66}, \bibinfo{number}{3} (\bibinfo{year}{2018}), \bibinfo{pages}{733--739}.
\newblock


\bibitem[Perold and Sharpe(1988)]%
        {perold1988dynamic}
\bibfield{author}{\bibinfo{person}{Andr{\'e}~F. Perold} {and} \bibinfo{person}{William~F. Sharpe}.} \bibinfo{year}{1988}\natexlab{}.
\newblock \showarticletitle{Dynamic Strategies for Asset Allocation}.
\newblock \bibinfo{journal}{\emph{Financial Analysts Journal}} \bibinfo{volume}{44}, \bibinfo{number}{1} (\bibinfo{year}{1988}), \bibinfo{pages}{16--27}.
\newblock


\bibitem[Schulman et~al\mbox{.}(2017)]%
        {schulman2017ppo}
\bibfield{author}{\bibinfo{person}{John Schulman}, \bibinfo{person}{Filip Wolski}, \bibinfo{person}{Prafulla Dhariwal}, \bibinfo{person}{Alec Radford}, {and} \bibinfo{person}{Oleg Klimov}.} \bibinfo{year}{2017}\natexlab{}.
\newblock \showarticletitle{Proximal Policy Optimization Algorithms}. In \bibinfo{booktitle}{\emph{ArXiv preprint arXiv:1707.06347}}.
\newblock


\bibitem[Sharpe(1964)]%
        {sharpe1964capital}
\bibfield{author}{\bibinfo{person}{William~F. Sharpe}.} \bibinfo{year}{1964}\natexlab{}.
\newblock \showarticletitle{Capital Asset Prices: A Theory of Market Equilibrium under Conditions of Risk}.
\newblock \bibinfo{journal}{\emph{The Journal of Finance}} \bibinfo{volume}{19}, \bibinfo{number}{3} (\bibinfo{year}{1964}), \bibinfo{pages}{425--442}.
\newblock


\bibitem[Vaswani et~al\mbox{.}(2017)]%
        {vaswani2017attention}
\bibfield{author}{\bibinfo{person}{Ashish Vaswani}, \bibinfo{person}{Noam Shazeer}, \bibinfo{person}{Niki Parmar}, \bibinfo{person}{Jakob Uszkoreit}, \bibinfo{person}{Llion Jones}, \bibinfo{person}{Aidan~N Gomez}, \bibinfo{person}{{\L}ukasz Kaiser}, {and} \bibinfo{person}{Illia Polosukhin}.} \bibinfo{year}{2017}\natexlab{}.
\newblock \showarticletitle{Attention Is All You Need}. In \bibinfo{booktitle}{\emph{Advances in Neural Information Processing Systems}}, Vol.~\bibinfo{volume}{30}. \bibinfo{pages}{5998--6008}.
\newblock


\bibitem[Wang et~al\mbox{.}(2021)]%
        {wang2021deeptrader}
\bibfield{author}{\bibinfo{person}{Zhicheng Wang}, \bibinfo{person}{Biwei Huang}, \bibinfo{person}{Shikui Tu}, \bibinfo{person}{Kun Zhang}, {and} \bibinfo{person}{Lei Xu}.} \bibinfo{year}{2021}\natexlab{}.
\newblock \showarticletitle{DeepTrader: A Deep Reinforcement Learning Approach for Risk-Return Balanced Portfolio Management with Market Conditions Embedding}. In \bibinfo{booktitle}{\emph{Proceedings of the AAAI Conference on Artificial Intelligence}}, Vol.~\bibinfo{volume}{35}. \bibinfo{publisher}{AAAI Press}, \bibinfo{pages}{643--650}.
\newblock
\href{https://doi.org/10.1609/aaai.v35i1.6164}{doi:\nolinkurl{10.1609/aaai.v35i1.6164}}


\bibitem[Wu et~al\mbox{.}(2021)]%
        {wu2021autoformer}
\bibfield{author}{\bibinfo{person}{Haixu Wu}, \bibinfo{person}{Jiehui Xu}, \bibinfo{person}{Jianmin Wang}, {and} \bibinfo{person}{Mingsheng Long}.} \bibinfo{year}{2021}\natexlab{}.
\newblock \showarticletitle{Autoformer: Decomposition transformers with auto-correlation for long-term series forecasting}.
\newblock \bibinfo{journal}{\emph{Advances in Neural Information Processing Systems}}  \bibinfo{volume}{34} (\bibinfo{year}{2021}), \bibinfo{pages}{22419--22430}.
\newblock


\bibitem[Xu et~al\mbox{.}(2020)]%
        {Xu2020}
\bibfield{author}{\bibinfo{person}{Ke Xu}, \bibinfo{person}{Yifan Zhang}, \bibinfo{person}{Deheng Ye}, \bibinfo{person}{Peilin Zhao}, {and} \bibinfo{person}{Mingkui Tan}.} \bibinfo{year}{2020}\natexlab{}.
\newblock \showarticletitle{Relation-Aware Transformer for Portfolio Policy Learning}. In \bibinfo{booktitle}{\emph{Proceedings of the 29th International Joint Conference on Artificial Intelligence (IJCAI-20)}}. \bibinfo{pages}{4647--4653}.
\newblock
\href{https://doi.org/10.24963/ijcai.2020/641}{doi:\nolinkurl{10.24963/ijcai.2020/641}}


\bibitem[Yan et~al\mbox{.}(2024)]%
        {Yan2024}
\bibfield{author}{\bibinfo{person}{Yunzhao Yan}, \bibinfo{person}{Ying Chen}, {and} \bibinfo{person}{Jian Wang}.} \bibinfo{year}{2024}\natexlab{}.
\newblock \showarticletitle{Deep Reinforcement Learning for Portfolio Optimization with Transaction Costs and Risk Control}.
\newblock \bibinfo{journal}{\emph{Quantitative Finance}} \bibinfo{volume}{24}, \bibinfo{number}{1} (\bibinfo{year}{2024}), \bibinfo{pages}{83--102}.
\newblock


\bibitem[Zhang and Yan(2023)]%
        {zhang2023crossformer}
\bibfield{author}{\bibinfo{person}{Yunhao Zhang} {and} \bibinfo{person}{Junchi Yan}.} \bibinfo{year}{2023}\natexlab{}.
\newblock \showarticletitle{Crossformer: Transformer Utilizing Cross-Dimension Dependency for Multivariate Time Series Forecasting}. In \bibinfo{booktitle}{\emph{International Conference on Learning Representations}}.
\newblock


\bibitem[Zhang et~al\mbox{.}(2020)]%
        {zhang2020cost}
\bibfield{author}{\bibinfo{person}{Yifan Zhang}, \bibinfo{person}{Peilin Zhao}, \bibinfo{person}{Qingyao Wu}, \bibinfo{person}{Bin Li}, \bibinfo{person}{Junzhou Huang}, {and} \bibinfo{person}{Mingkui Tan}.} \bibinfo{year}{2020}\natexlab{}.
\newblock \showarticletitle{Cost-Sensitive Portfolio Selection via Deep Reinforcement Learning}.
\newblock \bibinfo{journal}{\emph{arXiv preprint arXiv:2003.03051}} (\bibinfo{year}{2020}).
\newblock


\bibitem[Zhang et~al\mbox{.}(2022)]%
        {Zhang2022}
\bibfield{author}{\bibinfo{person}{Yifan Zhang}, \bibinfo{person}{Peilin Zhao}, \bibinfo{person}{Qingyao Wu}, \bibinfo{person}{Bin Li}, \bibinfo{person}{Junzhou Huang}, {and} \bibinfo{person}{Mingkui Tan}.} \bibinfo{year}{2022}\natexlab{}.
\newblock \showarticletitle{Cost-sensitive portfolio selection via deep reinforcement learning}.
\newblock \bibinfo{journal}{\emph{IEEE Transactions on Knowledge and Data Engineering}} \bibinfo{volume}{34}, \bibinfo{number}{1} (\bibinfo{year}{2022}), \bibinfo{pages}{236--248}.
\newblock
\href{https://doi.org/10.1109/TKDE.2020.2988572}{doi:\nolinkurl{10.1109/TKDE.2020.2988572}}


\bibitem[Zhou et~al\mbox{.}(2021)]%
        {zhou2021informerefficienttransformerlong}
\bibfield{author}{\bibinfo{person}{Haoyi Zhou}, \bibinfo{person}{Shanghang Zhang}, \bibinfo{person}{Jieqi Peng}, \bibinfo{person}{Shuai Zhang}, \bibinfo{person}{Jianxin Li}, \bibinfo{person}{Hui Xiong}, {and} \bibinfo{person}{Wancai Zhang}.} \bibinfo{year}{2021}\natexlab{}.
\newblock \bibinfo{title}{Informer: Beyond Efficient Transformer for Long Sequence Time-Series Forecasting}.
\newblock
\showeprint[arxiv]{2012.07436}~[cs.LG]
\urldef\tempurl%
\url{https://arxiv.org/abs/2012.07436}
\showURL{%
\tempurl}


\bibitem[Zhou et~al\mbox{.}(2022)]%
        {zhou2022fedformer}
\bibfield{author}{\bibinfo{person}{Tian Zhou}, \bibinfo{person}{Ziqing Ma}, \bibinfo{person}{Qingsong Wen}, \bibinfo{person}{Xue Wang}, \bibinfo{person}{Liang Sun}, {and} \bibinfo{person}{Rong Jin}.} \bibinfo{year}{2022}\natexlab{}.
\newblock \showarticletitle{Fedformer: Frequency Enhanced Decomposed Transformer for Long-Term Series Forecasting}. In \bibinfo{booktitle}{\emph{International Conference on Machine Learning}}. PMLR, \bibinfo{pages}{27268--27286}.
\newblock


\end{thebibliography}











\end{document}